\documentstyle[11pt,conf_iap]{article}
\begin{document}
\heading{GALAXY VELOCITY FIELD IN \\ TILTED COLD DARK MATTER MODELS}

\author{F. LUCCHIN$^{1}$, S. MATARRESE$^{2}$, L. MOSCARDINI$^{1}$, G.
       TORMEN$^{1}$}
       {$^{1}$ Dipartimento di Astronomia, Universit\`a di Padova,
       vicolo dell'Osservatorio 5, 35122 Padova, Italy.}
       {$^{2}$ Dipartimento di Fisica G. Galilei, Universit\`a di Padova,
       via Marzolo 8, 35131 Padova, Italy.}

\begin{abstract}{\baselineskip 0.4cm
We study the cosmic peculiar velocity field as traced by a sample of 1184
spiral, elliptical and S0 galaxies, grouped in 704 objects. We carry out a
statistical analysis, by calculating bulk flows and velocity correlation
functions for this sample and for mock catalogs which we extract from N--body
simulations. For the simulations we consider tilted (i.e. with spectral index
$n\leq 1$) CDM models with different values of the linear bias parameter $b$.
By mean of a maximum likelihood analysis we estimate the ability of the models
in fitting the observations as measured by the above statistics and in
reproducing the Local Group properties.}
\end{abstract}

\section{Introduction}
The Standard Cold Dark Matter (SCDM) scenario for structure formation possesses
a high predictive power and can explain many observed properties of the
large--scale structure of the universe. SCDM is characterized by a primordial
scale--invariant spectrum, $P(k) \propto k^n$, with spectral index $n=1$, of
Gaussian adiabatic perturbations in an Einstein--de Sitter universe and
vanishing cosmological constant. It has became usual to parametrize the
amplitude of the primordial perturbations by mean of the linear {\it bias}
parameter $b$, defined as the inverse of the {\it rms} mass fluctuation on a
scale of $R_8\equiv 8~h^{-1}$ Mpc:
\begin{equation}
b^2 \equiv  1/\sigma^2(R_8) = 2\pi^2 \left/ \int_0^\infty dk \right.
{}~k^2 ~P(k) ~W^2_{TH}(kR_8),
\end{equation}
where $W_{TH}(kR)=(3/kR) j_1(kR)$ is the top--hat window function and $j_\ell$
denotes the $\ell$--th order spherical Bessel function. We adopt the value
$h=0.5$ for the Hubble constant $H_0$ in units of $100$ km ${\rm s^{-1}}$.
The COBE
detection \cite{SMO2} of large angular scale Cosmic Microwave Background (CMB)
anisotropies can be used to normalize the CDM power--spectrum, resulting in $b
\approx 0.8$. However, the SCDM model has met increasing problems, mostly due
to the high ratio of small to large--scale power: in particular the model with
the COBE normalization predicts excessive velocity dispersion on scales of
order $1~h^{-1}$ Mpc and is unable to reproduce the slope of the galaxy angular
correlation function obtained in the APM survey (\cite{MAD}; see, however,
\cite{FON}).

A well known natural way to reduce the high ratio of small to large--scale
power of the SCDM model is to ``tilt" the spectral index of primordial
perturbations: ``tilted", i.e. $n<1$, CDM (hereafter TCDM) models boost power
from small to large scales (e.g. \cite{VIT, TLM, ADA, CEN, TOR}). The COBE DMR
experiment has lead to renewed interest in these models: the observed
anisotropy is in fact consistent with $n=1.15^{+.45}_{-.65}$ on scales $\geq
10^{3}~h^{-1}$ Mpc. It is evident that the COBE normalization of TCDM models
imply reduced power on all scales below $10^3$ Mpc. Moreover a large number of
post--COBE papers (see \cite{CRI} and references therein) pointed out that
properly accounting for the gravitational--wave contribution to the
Sachs--Wolfe effect leads to a relevant enhancement of the linear biasing
factor. This effect is relevant in power--law inflation: \cite{LMM} obtained
\begin{equation}
b(n) \approx 0.80 {\sqrt {14-12n \over 3-n}}
\times 10^{1.20(1-n)} (1 \pm 0.17).
\end{equation}
\\
In a recent work \cite{TOR} we analyzed the peculiar velocity field traced by
optically selected galaxies, which probes the primordial spectrum up to scales
$\sim 100~h^{-1}$ Mpc. The results were then compared with similar analyses
carried out on mock catalogs in Monte Carlo simulations obtained from linear
theory in $n\leq 1$, $\Omega_0 \leq 1$ CDM models, for different values of the
bias factor, by assuming that the galaxy velocity field gives an unbiased
signal of the underlying mass distribution. We report here the preliminary
results of a forthcoming paper \cite{MOS} where the same analysis is
performed on
mock catalogs extracted from N--body simulations in $\Omega = 1$ models. A
similar method has been applied also to simulations with skewed (i.e.
non--Gaussian) CDM initial conditions: first results are presented in
\cite{NG}.

\section{The data}
The sample we used was compiled from the ``Mark II" data files kindly supplied
by David Burstein, a collection of several data, including distances and
peculiar velocities, for more than one thousand spirals, ellipticals and S0
galaxies. The sample include:

\noindent
i) the Aaronson `good' and `fair' field spirals \cite{FB} and the de
Vaucouleurs and Peters spirals \cite{DEV};

\noindent
ii) the Aaronson cluster spirals \cite{AAB,AAR};

\noindent
iii) the ellipticals and S0, which combine the survey by \cite{LYN} with the
data by \cite{LUC} and \cite{DRE}.

\noindent
The galaxies were grouped following the rules in \cite{LYN,FAB}; we also
considered every Aaronson cluster of spirals as a single object. The procedure
reduces distance uncertainties of a factor ${\sqrt N}$, if $N$ is the number of
grouped galaxies. Our final big  sample  consists at the end of 1184 galaxies
grouped in 704 objects. In Table 1 we give a summary of the different
subsamples, for each one indicating the number of galaxies and grouped objects.
\medskip
\begin{center}
{\bf Table 1.} Samples.
\end{center}
\begin{center}
\begin{tabular}{|l|l|l|}
\hline
\hskip1.8cm Sample & Number of Galaxies & Number of Objects \\
\hline
Aaronson `good' spirals & \hskip1.3cm ~224  & \hskip1.3cm 224 \\
Aaronson `fair' spirals & \hskip1.3cm ~139  & \hskip1.3cm 139 \\
Aaronson cluster spirals & \hskip1.3cm ~204  & \hskip1.3cm ~17 \\
de Vaucouleurs--Peter spirals & \hskip1.3cm ~~73  & \hskip1.3cm ~73 \\
Elliptical galaxies & \hskip1.3cm ~544 & \hskip1.3cm 251 \\
Total sample & \hskip1.3cm 1184 & \hskip1.3cm 704 \\
\hline
\end{tabular}
\end{center}
\medskip
All distances in the sample have a uniform Malmquist bias correction. Residual
Malmquist bias due to clustered structures still affects the data, but removing
it would require the knowledge of their selection function, which is not our
case, because the spiral subsamples do not have a clear selection criterion
(see, e.g. \cite{DEK}).
Nevertheless, the grouping procedure helps in reducing the
residual bias, besides reflecting the fact that different galaxies belonging to
the same group or cluster actually map only one point of the peculiar velocity
field.

\section{Statistical analysis of the velocity field}
The {\it velocity dipole}
(or {\it bulk flow}) for a galaxy catalog with $N$ objects, endowed with
peculiar velocities ${\bf v}_i$, is defined through a least--squares fit (e.g.
\cite{REG})
\begin{equation}
v_{bulk}^\alpha = (M^{-1})^{\alpha\beta} \sum_{i=1}^N u_i^\beta
\end{equation}
(summation over repeated indices is understood); $u_i^\alpha \equiv ({\bf v}_i
\cdot {\hat {\bf r}}_i) {\hat {\bf r}}_i^\alpha$ is the $\alpha$ component
($\alpha=1,2,3$) of the radial peculiar velocity of the $i$--th galaxy. The
{\it projection matrix}
\begin{equation}
M^{\alpha\beta} = \sum_{i=1}^N \hat
r_i^\alpha \hat r_i^\beta
\end{equation}
takes into account the geometry of the sample.
\\
The second statistics we considered is the {\it velocity correlation function}.
Various alternative have been proposed in the literature; we chose the one
proposed by \cite{GOR}, which reads
\begin{equation}
\psi_1(r) = \sum_{pairs(r)} {\bf u}_1 \cdot {\bf u}_2 \left/
\sum_{pairs(r)} \hat {\bf r}_1 \cdot \hat {\bf r}_2,\right.
\end{equation}
where the sum is extended over galaxy pairs separated by a distance $r$.

\section{N--body simulations and mock catalogs}
In order to mimic the large--scale peculiar velocity field we performed two
N--body simulations using a particle--mesh code with $N_p = 128^3$ particles on
$N_g = 128^3$ grid points; the box--size was $260~h^{-1}$ Mpc. The velocity
field was assumed to be Gaussian with power--spectrum
\begin{equation}
P_v(k) \propto  k^{n-2} T^2(k),
\end{equation}
where $T(k)$ is the CDM transfer function \cite{DEFW},
\begin{equation}
T(k) = [1 + 6.8 k + 72.0 k^{3/2} + 16.0 k^2 ]^{-1}.
\end{equation}
We ran two simulations of 12 models, combinations of the values $n=0.6$, 0.8, 1
and $b= 1$, 1.5, 2, 2.5. We will show below the results only for three basic
models: $(n,~b) = (0.6,~2.5)$ (high tilt and bias); $(n,~b) = (0.8,~1.5)$
(moderate tilt and bias); $(n,~b) = (1,~1)$ (no tilt and no bias).

We defined the velocity field of each simulation by interpolating the
particles' velocities onto a cubic grid with $128^3$ grid points, using
a TSC algorithm and a further Gaussian smoothing with filter width $275$
km ${\rm s^{-1}}$.

We built up our simulated catalogs (e.g., \cite{GOR,DSY}) by locating for each
simulation 25 ``observers" in grid--points with features similar to those of
the Local Group (LG). The requirements are the following:

\noindent
i) the peculiar velocity $v$ is in the range of the measured LG motion,
$v_{LG,obs} = 622 \pm 20$ km ${\rm s^{-1}}$;

\noindent
ii) the local flow is quiet, i.e. that the local `shear' is small, ${\cal S}
\equiv |{\bf v} - \langle {\bf v} \rangle|/| {\bf v}| < 0.5$, where $\langle
{\bf v} \rangle$ is the average velocity of a sphere of radius
$R= 750$ km ${\rm s^{-1}}$
centered on the LG;

\noindent
iii) the density contrast in the same sphere is in the range $-0.2 < \delta <
1.0$.

\noindent
We then measured radial peculiar velocities by sampling the velocity field at
the same positions of the observed galaxies. The reference frame was fixed so
that the velocity vector at the central point singles out the CMB dipole
direction, while the direction of the remaining axis was selected at random.
\\
We take into account the random galaxy distance errors by perturbing each
distance and radial peculiar velocity with Gaussian noise (e.g. \cite{DEK}),
$r_{i,p} = r_i + \xi_i \Delta r_i $ and $u_{i,p} = u_i - \xi_i \Delta r_i +
\eta_i \sigma_f,$ where $\xi_i$ and $\eta_i$ are independent standard Gaussian
variables; $\Delta r_i$ is the estimated galaxy distance error and
$\sigma_f=200$ km ${\rm s^{-1}}$ is the Hubble flow noise.
Sampling the simulated velocity
field at the same positions of the observed galaxies introduces the same
sampling errors of the real data.

\section{Results}
We applied to our simulated catalogs a first statistic by characterizing the
observers using their velocity, local shear and local density contrast, as
discussed above. We found that the selection effect of these constraints
changed the ``observers" distribution. Consequently, the models
provided us with statistical results that turned out to be rather different
from simple unconstrained estimates. As expected, the constraints were more
effective for the models with observed features of the LG corresponding to
values far from the mean. The imposed constraints on the Local Group velocity
and shear exclude contributions from grid points with large velocity. This
translates in a reduced amplitude for the bulk motions and for the value of
$\Psi_1(r)$ at small separation when compared with equivalent unconstrained
simulations.
In Figure 1 we show for our basic models $(n,~b) = (0.6,~2.5), ~(0.8,~1.5),
{}~(1,~1)$ the probability distribution of the three quantities $v$, $\delta$
and ${\cal S}$. The vertical lines show the range allowed by the assumed Local
Group constraints.

\vspace{16.8cm}
\medskip
{\baselineskip 0.4cm
{\bf Figure 1.}
Probability distribution of the peculiar velocity $v$ (top row), density
contrast $\delta$ (central row) and local `shear' $\cal S$ (bottom row),
calculated on the grid--points from simulations of the models $(n,~b) =
(0.6,~2.5)$ (first column),  $(n,~b) = (0.8,~1.5)$ (central column) and $(n,~b)
= (1,~1)$ (last column). The vertical lines show the range allowed by the
different Local Group constraints.
}
\newpage

In Table 2 we report the percentage of grid points fulfilling each constraints
separately and altogether [${\cal P}(LG)$].
\medskip
\begin{center}
{\bf Table 2.} Local Group constraints.
\end{center}
\begin{center}
\begin{tabular}{|l|l|l|l|l|l|}
\hline
\hskip0.1cm $n$ & \hskip0.1cm $b$ &
${\cal P}(v)$ & ${\cal P}(\delta$) & ${\cal P}({\cal S}$) &
${\cal P}(LG)$ \\
\hline
0.6 & 2.5 & ~0.7 & 67.2 & 97.3 & ~~0.6 \\
0.8 & 1.5 & ~4.3 & 50.0 & 97.2 & ~~2.4 \\
1.0 & 1.0 & ~5.9 & 37.3 & 97.4 & ~~2.2 \\
\hline
\end{tabular}
\end{center}
\medskip
While the constraint on the local shear is poorly effective (no differences
between all the considered models), those on the density and velocity of the
Local Group turn out to depend mostly on the bias parameter and almost nothing
on the spectral index:
with higher values of $b$, higher ${\cal P}(\delta)$ and lower
${\cal P}(v)$ result, for all considered $n$. The total
probability ${\cal P}(LG)$ shows as best model $(n,~b) = (0.8,~1.5)$.

In Figure 2 we plot for the three typical models the bulk flow distribution
calculated from our mock catalogs. The continuous vertical line refers to the
observed value: we found for our composite galaxy sample $v_{bulk}=306 \pm 72$
km ${\rm s^{-1}}$, with a misalignment angle
$\alpha = 54^\circ \pm 13^\circ$ with respect
to the direction of the CMB dipole. The plotted observational errors take into
account the uncertainties due both to the sparse geometry (by bootstrap
resamplings of our catalog) and to the distance errors (by estimating the
dispersion after perturbing the true catalog with Gaussian errors).

\vspace{10.5cm}
\medskip
{\baselineskip 0.4cm
{\bf Figure 2.}
The probability distribution for the absolute value of the bulk flow,
$v_{bulk}$, for the models $(n,~b) = (0.6,~2.5), ~(0.8,~1.5), ~(1,~1)$
from the left to the right.
The vertical lines refers to the value obtained from our real catalog.
}

\bigskip\smallskip
\noindent
Due to the small number of mock catalogs, the probability distributions
are not well sampled and show an irregular behavior.
The preliminary results of this test
seem to indicate that models with high tilt and bias are
less unlikely than other models:
in fact only 14\% of the mock catalogs
obtained from the simulation $(n,~b) = (0.6,~2.5)$ present values
of $v_{bulk}$
in the observational range, while for $(n,~b) = (0.8,~1.5)$ and
$(n,~b) = (1,~1)$ the percentages of catalogs are 38\% and 42\% respectively.

Figure 3 compares the velocity correlation resulting from
our mock catalogs of the three typical models to the observed one.
We evaluated $\Psi_1(r)$ for the real data by
counting galaxy pairs in separation bins of $500$ km ${\rm s^{-1}}$ up
to a maximum
separation of $5,000$ km ${\rm s^{-1}}$.
The error bars were estimated as for bulk
flow and take into account both the sparse sampling of the data and
the distance
errors.

\vspace{10.5cm}
\medskip
{\baselineskip 0.4cm
{\bf Figure 3.}
The observed velocity correlation function vs. the separation $r$ (thick solid
line with squares; error bars are one standard deviation for each bin) compared
to the probability distribution for $\Psi_1$ from the simulated catalogs. Left
panel: high tilt and bias. Central panel: moderate tilt and bias. Right panel:
no tilt and no bias. The different lines refer to the $5\%$, $25\%$, $50\%$,
$75\%$ and $95\%$ percentiles.}

\bigskip\bigskip
The simulated distributions are very different: models with high tilt and
bias present a narrow distribution. The widest distribution is for
$(n,~b) = (1,~1)$.
\\
In order to compare observations vs. models, we choose to use,
between the different possible statistics
discussed in \cite{TOR}, the integral of $\Psi_1(r)$
from zero to the maximum considered separation of pairs,
$R_{max}=5,000$ km ${\rm s^{-1}}$
(see also \cite{GOR}):
\begin{equation}
J_v = \int_0^{R_{max}} \Psi_1(r) \ dr.
\end{equation}
This is a simple
one--dimensional statistic, bearing as much information as possible on the
original correlation function without the necessity to have a
very large number of simulated catalogs to sample the whole distribution
of $\Psi_1$.
In fact the velocity correlation function is a random function of
the separation $r$, which can assume infinite values; its probability
distribution is then a functional, or at least an $N$--variate distribution
if we sample this function with $N$ bins (10 in our case).

We calculated for our typical models the percentage of the simulated catalogs
whose value of $J_v$ is similar (i.e. less than one standard deviation
different) to the observed one . We found the percentages
28\%, 48\% and 64\% for $(n,~b) = (0.6,~2.5), ~(0.8,~1.5), ~(1,~1)$
respectively.
As general trend we can state that low values of the bias parameter are
preferred, in particular in connection with low tilt.

\section{Maximum likelihood}

We performed a maximum likelihood analysis to compare the statistics from
different simulations.
Calling $\vec A$ the random vector of the statistics
we used to constrain the simulated LG, $\vec A=(v_{LG}, {\cal S}, \delta)$, and
$\vec B$ the vector of all other statistics,
$\vec B=(v_{bulk}, \alpha, J_v)$, the joint probability
distribution of $\vec A$ and $\vec B$, under the condition
$\vec A=\vec A_{obs}$, is ${\cal P}(\vec A_{obs},\vec B)=
{\cal P}(\vec A_{obs}) {\cal P}(\vec B |\vec A_{obs})$.
For a given model $H$, the likelihood function is
${\cal L}(H)={\cal P}(\vec A_{obs}|H) {\cal P}(\vec B_{obs}|\vec A_{obs},H)$;
since we considered models that differ by their values of $n$ and
$b$, ${\cal L}={\cal L}(n,~b)$.
The joint likelihood ${\cal P}(\vec B_{obs}|\vec A_{obs},H)$
of $v_{bulk}$, misalignment angle $\alpha$ and $J_v$
has been computed counting the number of simulated catalogs that
have, at the same time, bulk flow, $\alpha$ and
$J_v$ equal to the observed ones,
up to the
fixed tolerance (i.e. one observational $\sigma$).

In the case of our three typical models we found
0.01\%, 0.08\% and 0.17\% for the models $(n,~b) =
(0.6,~2.5), ~(0.8,~1.5), ~(1,~1)$ respectively.
If we extend our analysis to all 12 original models, we found that the
best model is $(n,~b) = (0.8,~1)$, but the likelihood is very flat
in the region $(n,~b) = (0.8-1,~1-1.5)$ and differences between models
inside this region are not significative.
Using a {\it Chi--square} approximation to provide confidence levels
to our predictions,
we can reject, in any case,  the model with high bias and tilt
$(n,~b) = (0.6,~2.5)$ at the $90\%$
confidence level.

\section{Conclusions}

In this paper we report the first results from a statistical analysis of the
large--scale velocity field in the context of tilted CDM models. We extend our
previous work based on Monte Carlo simulations in linear theory \cite{TOR}
using N--body simulations, i.e. in strongly non--linear regime. We consider  12
models, combinations of the values 0.6, 0.8, 1 for the spectral index $n$ and
1, 1.5, 2, 2.5 for the bias parameter $b$. We calculate the probability to have
grid points with features similar to Local Group, bulk flows and velocity
correlation functions for mock galaxy catalogs and compare the resulting
distributions with the results of a composite sample of 1184 galaxies, grouped
in 704 objects. Using a maximum likelihood method we calculate the probability
of the models to reproduce the observations as measured by the above
statistics.

Our results, even if obtained from a small number of mock catalogs and
consequently affected by a larger statistical uncertainty, essentially confirm
those derived from Monte Carlo simulations \cite{TOR}.

In particular model with high tilt ($n \le 0.6$) are rejected by the
combination of the COBE results \cite{SMO2} and the present analysis. The best
model is $(n,~b)= (0.8,~1)$, but the likelihood function is
nearly flat in the ranges $0.8 \le n \le 1$ and $1 \le b \le 1.5$. Note that
COBE implies for $n=0.8$ and $b=1.35\pm0.3$ a negligible contribution
from gravitational waves; in any case
the constraints on small--scale velocity dispersion
prefer $ n < 1$ and/or $ b > 1$.

Moreover our results show that lower values of the bias parameter
are preferred; this implies that the gravitational waves
contribution to $\Delta T/T$ at large scales should be negligible.

As a general result, our more accurate treatment of the errors in
the observational data shows
that tilted CDM models are not excluded by the combination of COBE and
present analysis. Of course, the possibility of having larger sample of data
(e.g. ``Mark III", see \cite{TB}) can help to increase the discriminatory
power of these statistical tests on the large--scale velocity field.

\acknowledgements{
We thank Antonio Messina for technical help.
This work has been partially supported by Ministero dell'Universit\`a
e della Ricerca Scientifica e Tecnologica. The staff and the management
of the CINECA Computer Center
(Bologna) are warmly acknowledged for their assistance and for allowing the
use of computational facilities.}

\vfill
\end{document}